\documentclass[prb,twocolumn,showpacs,amsmath,amssymb,superscriptaddress]{revtex4}

\usepackage{graphicx}
\usepackage{dcolumn}
\usepackage{bm}
\usepackage{amsmath}

\begin{document}

\preprint{APS/123-QED}

\title{\textit{Ab initio} quality neural-network potential for sodium}

\author{Hagai Eshet}
\email{hagai.eshet@gmail.com}
\author{Rustam Z. Khaliullin}
\email{rustam@khaliullin.com}
\affiliation{
Department of Chemistry and Applied Biosciences, ETH Z\"urich, USI Campus, via G. Buffi 13, 6900 Lugano, Switzerland
}
\author{Thomas D. K\"uhne}
\affiliation{
Institute of Physical Chemistry and Center of Computational Sciences, Staudinger Weg 9, D-55128 Mainz, Germany
}
\affiliation{
Johannes Gutenberg University of Mainz, Staudinger Weg 9, D-55128 Mainz, Germany
}
\author{J\"org Behler}
\affiliation{
Lehrstuhl f\"ur Theoretische Chemie, Ruhr-Universit\"at Bochum, D-44780 Bochum, Germany
}
\author{Michele Parrinello}
\affiliation{
Department of Chemistry and Applied Biosciences, ETH Z\"urich, USI Campus, via G. Buffi 13, 6900 Lugano, Switzerland
}

\date{\today}

\begin{abstract}
An interatomic potential for high-pressure high-temperature (HPHT) crystalline and liquid phases of sodium is created using a neural-­network (NN) representation of the \textit{ab initio} potential energy surface. It is demonstrated that the NN potential provides an \textit{ab initio} quality description of multiple properties of liquid sodium and bcc, fcc, cI16 crystal phases in the P--T region up to 120~GPa and 1200~K. The unique combination of computational efficiency of the NN potential and its ability to reproduce quantitatively experimental properties of sodium in the wide P--T range enables molecular dynamics simulations of physicochemical processes in HPHT sodium of unprecedented quality.
\end{abstract}

\maketitle

\section{Introduction}

\begin{figure}
\includegraphics*[width=8.5cm]{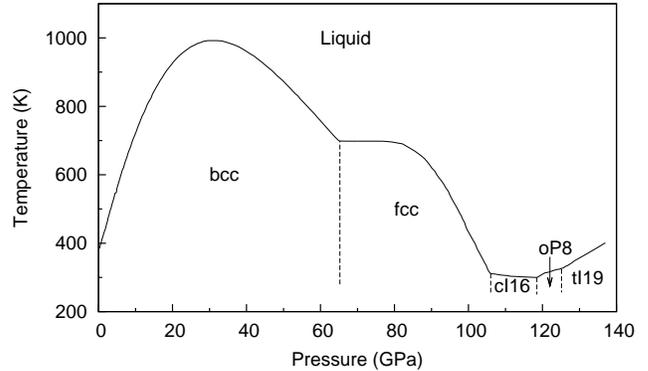}
\caption{\label{fig:phd} Experimental phase diagram of sodium~\cite{a:gregorprl}.}
\end{figure}

Recent experimental studies have shown that sodium, a simple metal at ambient conditions, exhibits unexpected complex behavior under high pressure (FIG.~\ref{fig:phd}). At ambient conditions, the thermodynamically stable form of Na is the highly-symmetric bcc structure, which transforms upon compression to the fcc phase at 65$\pm$1~GPa~\cite{a:hanfland} and then to a more complex cI16 structure at 105$\pm$1~GPa~\cite{a:syassen,a:mcmahon}. Above 118~GPa, sodium adopts a primitive orthorhombic structure with eight atoms per unit cell (oP8), which transforms at 125$\pm$2~GPa to an incommensurate composite host-guest structure (tI19)~\cite{a:gregor}. It has been reported recently that, at pressure around 118$\pm$2~GPa, sodium crystallizes in a number of very complex low-symmetry crystal structures with 50 to 512 atoms in the unit cell~\cite{a:gregor}. Very recently, a new optically transparent double-hexagonal closed packed phase of Na (hP4) has been observed above 200~GPa~\cite{a:na-oganov}. The behavior of the sodium melting line has also been discovered to be unusual. Measurements have revealed an unprecedented pressure-induced drop in melting temperature from 1,000~K at $\sim$30~GPa to room temperature at $\sim$120~GPa~\cite{a:gregorprl,a:raty}.

The complexity of the sodium phase diagram combined with experimental difficulties of obtaining detailed characterization of HPHT phases~\cite{a:gregorprl} make molecular dynamics (MD) simulations an indispensible tool for the investigation of the microscopic origins of complex behavior in dense sodium. The reliability of MD simulations depends crucially on the quality of the underlying potential energy surface (PES). While density functional theory (DFT) provides a comprehensive framework for modeling a wide variety of sodium structures it is not practical for lengthy MD simulations because of its high computational cost. On the other hand, the construction of accurate and computationally efficient potentials capa­ble of describing various bonding patterns in HPHT sodium is a formidable challenge. Many potentials for sodium devel­oped based on the pseudopotential theory~\cite{a:magana,a:hu6,a:hu7} and the embedded atom model (EAM)~\cite{a:hu10,a:hu11,a:hu13,a:hu14,a:hu15,a:hu,a:belashchenko} are limited to a few phases in a narrow P--T region of the phase diagram and do not always give a correct description of all properties or phenomena of interest. 

In this paper, we present an interatomic potential for sodium based on a recently developed high-dimensional neural-network (NN) representation of \textit{ab initio} PESs~\cite{a:behler}. In this approach, the sodium PES is represented by a highly flexible NN optimized to reproduce high-quality DFT energies of a large dataset of sodium structures. The dataset includes crystal (bcc, fcc, cI16, oP8) and liquid structures for pressures up to 120~GPa and temperatures up to 1200~K. We demonstrate that the NN potential is capable of reproducing numerous properties of sodium phases in this P--T region with an accuracy comparable with that of the underlying DFT calculations. From a computational standpoint, the NN energies, forces, and stress tensor are evaluated with the speed of empirical potentials enabling \textit{ab initio} quality MD simulations of sodium on unprecedented length and time scales. 

\section{Computational methods}

\subsection{NN representation of \textit{ab initio} PESs}

Artificial neural networks (NNs), biology-inspired machine learning algorithms, are emerging as a new class of interatomic potentials that combine the accuracy of an \textit{ab initio} description of PESs with the efficiency of empirical force fields~\cite{a:nnblank,a:nnprudente,a:nnlorenz,a:nnscheffler,a:nnmanzhos,a:nnreuter,a:behler,a:nnraff1,a:behler3}. In this class, the PES is represented by a highly flexible NN capable of describing various bonding patterns in the system. Given a number of atomic configurations for which the \textit{ab initio} energies are known the NN is tuned to reproduce these energies in the training process. Overfitting (i.e. obtaining a good fit to the training data, but performing less accurately when making predictions) is controlled by testing the performance of the NN for an independent test set not used in the optimization. Once trained, the NN performs interpolation to reconstruct the potential energy for new atomic configurations with the speed of empirical potentials and is, thus, useful to perform lengthy MD runs. 

The NN methodology eliminates many problems associated with empirical potentials. First and foremost, NNs completely obviate the problem of guessing a complicated functional form of the interatomic potential. This form is determined automatically by the NN. Second, the entire training procedure is fully automated so that NNs can be readily extended to new regions of the PES. Thus, the significant human effort normally required to (re)parametrize the potential is replaced with a short computer calculation. Finally, the accurate mapping of \textit{ab initio} energies ensures that \emph{all} properties determined by the topology of the PES are described with the accuracy comparable with that of first-principle calculations.

Neural networks have been successfully used to in­terpolate PESs of simple chemical systems for the last
decade~\cite{a:nnscheffler,a:nnmanzhos,a:nnprudente,a:nnraff1}. However, NN-­based potentials that can be used to map high-dimensional PESs of bulk systems are very rare~\cite{a:behler,a:behler3,a:sanville}. Here, we used the NN methodology introduced recently by Behler and Parrinello~\cite{a:behler}. In this NN scheme, the total energy of the system is expressed as a sum of atomic energy contributions. The atomic energies are calculated by a standard feedforward NN as a function of the energetically relevant local geometric environment of each atom. The local environment of a given atom is described by several order parameters called symmetry functions, which include radial and angular many-body terms and depend on the positions of all neighbors within a specified cutoff radius. The use of symmetry functions (instead of cartesian coordinates) as NN input(s) and the partitioning of the total energy into atomic contributions ensures that all quantities computed with the NN (energies, analytical forces, and stress tensors) are invariant to translations, rotations, and the order of atoms. Futhermore, once the fit is obtained, such an NN potential can be applied to systems containing an arbitrary number of atoms.

Details of the high-dimensional NN method are given elsewhere~\cite{a:behler,a:behler3}. Several recent works have demonstrated applicability of this methodology to modeling phase diagrams~\cite{a:khalcoex} as well as structures of liquids~\cite{a:behler} and crystals~\cite{a:behler2,a:behler3}. 

%

\subsection{NN potential for HPHT sodium}

The accuracy of the reference \textit{ab initio} energies is of paramount importance while training the NN. We employed the PBE functional in combination with an ultrasoft pseudopotential with the $2s$ and $2p$ semicore electrons included explicitly as the valence states.  A large plane-wave cutoff of 100~Ry and a dense mesh of $k$-­points ($22\times 22\times 22$ for the primitive cell of bcc and fcc, $12\times 12\times 12$ for the primitive cell of cI16 and oP8, and 6000\textbf{k} for liquid) were used for all structures so as to ensure convergence of the total energy to 1.5~meV/atom. The Quantum-Espresso package~\cite{a:pwscf} was used to perform all \textit{ab initio} calculations.

The initial fitting of the sodium NN potential was performed on crystal structures that included the zero­-temperature and randomly distorted structures of bcc, fcc, cI16, and oP8 phases in the pressure range from -1 to 200~GPa. Liquid structures of sodium were modeled with periodic cubic cells containing 32 and 64 randomly arranged atoms with the densities corresponding to the 0--120~GPa pressure range. The energetically relevant local environment for each atom is defined by 48 order parameters (see Ref.~\onlinecite{a:behler3} for details) constructed to include all neighbors within 6.0\AA ~cutoff radius.

After the initial training, the NN was improved self­-consistently by iterative repetition of the NN-­driven MD and metadynamics-accelerated Parrinello-Rahman simulations~\cite{a:rahman}, collection of new structures emerging from the simulations, calculation of the DFT energies for the physically relevant structures, and refinement of the NN. These iterations were performed until the root mean square error (RMSE) of the new structures not included in the fit converged to the RMSE of the test set. After the self-­consistent procedure, the DFT dataset contained $\sim$17,000 DFT energies corresponding to more than 350,000 atomic environments. 10\% of all structures were ran­domly chosen for the test set. The best fit was ob­tained for a NN with 2 hidden layers, each of which contains 25 nodes (the total number of the NN parameters is 1901). The RMSE of the training set is 0.72~meV/atom, while the RMSE of the test set is 0.91~meV/atom. The maximum absolute errors are 6.18~meV/atom and 7.62~meV/atom for the training and test sets, respectively. We would like to emphasize that the fitting procedure introduces only small error (less than 1~meV/atom) in addition to the 1.5~meV/atom numerical (convergence) error of the DFT calculations. Thus, the NN-potential for sodium is expected to reproduce closely the \textit{ab initio} PES.


\begin{figure}
\includegraphics*[width=8.5cm]{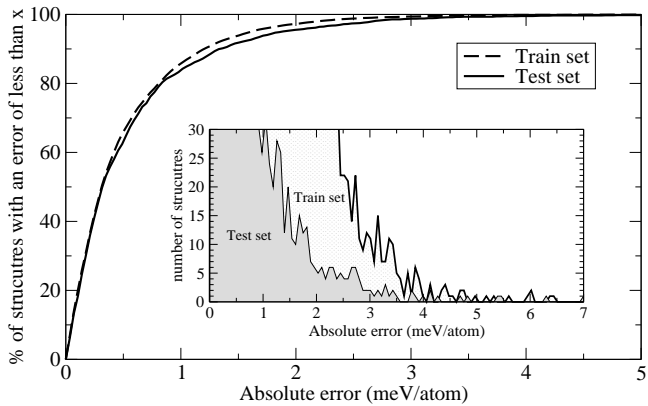}
\caption{\label{fig:errors} Normalized cumulative histogram of the absolute NN errors in the training and test sets. It demonstrates that the fitting error is less than 1~meV/atom for 83\% of structures and less than 2~meV/atom for 95\% of structures in the test set. The inset shows the ordinary histogram for the same data. Only a few strucutres in the test and train sets have errors larger than 3~meV/atom.}
\end{figure}

\subsection{\label{ssec:compd} Simulation details}

The zero-temperature structures of sodium crystals were obtained by the minimization of the enthalpy with respect to the atomic coordinates and lattice parameters at constant (external) pressure. The zero-pressure zero-temperature bulk moduli and their pressure derivatives were calculated by fitting the energy to the Murnaghan equation~\cite{a:murnaghan}. The second­-order elastic constants were determined from a fit of the energy as a function of an appropriate cell distortion to a parabola~\cite{a:guo}.

The lattice dynamics calculations were performed using the linear response method within the density functional perturbation theory~\cite{a:testa}. An $8\times 8\times 8$ $q$-mesh was used in the interpolation of the force constants for the phonon dispersion curve calculations. The NN lattice dynamics calulations were carried out with the direct supercell calculation method~\cite{a:parlinski}. Convergence tests suggested the use of a $6\times 6\times 6$ supercell in the force constant calculation. The PHON program~\cite{a:phon} was used to obtain the phonon dispersion curves.

The radial distribution functions $g(r)$, thermal expansion coefficients, isothermal compressibility, self-diffusion and viscosity coefficients for liquid sodium were obtained from NN-driven MD simulations. All MD runs were performed with the DLPOLY package~\cite{a:dlpoly} interfaced with the NN code. The temperature was controlled using a colored­-noise Langevin thermostat that was tuned to provide the optimum sampling efficiency over all relevant vibrational modes~\cite{a:ceriotti}. Constant-pressure simulations were governed by Nos\'e--Hoover equations of motion with Langevin noise on the particle and cell velocities~\cite{a:feller,a:ceriotti}. The time step was set to 1.0~fs. 

State points along several isotherms ($T$ from 400 to 1000~K with 100~K increment) were obtained from NPT simulations with cells of 512 atoms. The density at each $P$ and $T$ was obtained by averaging over a 25~ps trajectory. The volumetric thermal expansion coefficients do not change appreciably with temperature and were evaluated for the 800--1000~K range assuming a linear density dependence on temperature. The radial distribution functions were obtained from 25~ps NVT trajectories for systems of 54 atoms.

To evaluate dynamical properties of a liquid one must address the issue of size dependence of the self-diffusion coefficient. This effect has been analyzed by D\"unweg and Kremer~\cite{a:thomas63,a:thomas64} who established the following dependence for the apparent diffusion coefficient $D$ on the simulation box length $L$:
\begin{equation}\label{eq:diff}
D(L)=D(\infty)-\frac{2.837297 k_B T}{6 \pi \eta L},
\end{equation}
where $D(\infty)$ is the true diffusion coefficient and $\eta$ is the translational shear viscosity, which is much less system size dependent than $D$~\cite{a:kuhnewater}. To calculate $D(\infty)$ and $\eta$, apparent diffusion coefficients were computed for different system sizes. $D(\infty)$ and $\eta$ were then obtained from the y-intercept and the slope, respectively, of a linear fit of $D(L)$ with respect to $1/L$.

\begin{figure}
\includegraphics*[width=8.5cm]{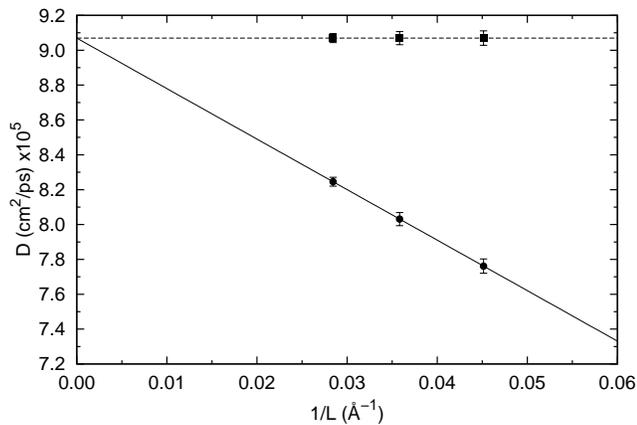}
\caption{\label{fig:dofl} Diffusion coefficient at $T=490$~K as a function of the inverse box length. Circles show apparent diffusion coefficients with the linear fit represented by the solid line. Squares show the diffusion coefficients $D(\infty)$ corrected for finite-size effects with Eq.~\ref{eq:diff}. The constant dashed line marks the value of $D(\infty)$.}
\end{figure}

It is important to emphasize that long MD trajectories are essential to obtain statistically accurate results for transport properties of liquids. Furthermore, it is desirable to perform simulations using large systems. Hence, di­rect \textit{ab initio} MD simulations (especially with a large plane wave cutoff and a dense $k$-­point mesh) are computationally very demanding for the evaluation of the diffusion and viscosity coefficients~\cite{a:kuhnewater}, whereas, the NN provides an affordable and accurate method to determine $D(\infty)$ and $\eta$. Each apparent diffusion coefficient was calculated from 10 independent 500-ps NVE trajectories. Systems containing 256, 512, and 1024 atoms were used to determine the dependence of $D$ on $L$ (Figure~\ref{fig:dofl}). Thus, the total simulation time required to obtain $D(\infty)$ and $\eta$ at four temperature points is $\sim$60~ns, which clearly demonstrates the advantage of the NN approach in comparison with direct \textit{ab initio} simulations.

\section{Results and discussion}

\subsection{Solid phases}

The first test of the NN potential was to calculate structural, elastic, and vibrational properties for the zero-temperature structures of the two most important crystal phases of sodium, bcc and fcc. The computed zero-pressure values for the latice constants and stiffness coefficients are summa­rized and compared with DFT and experimental values in TABLE~\ref{tab:zerok}. The NN accurately reproduces zero-temperature zero-pressure DFT values for lattice constants (the error does not exceed 0.02\%) and all independent elastic constants (the error is less than 5\% for $c_{11}$, $c_{12}$, $c_{44}$). The pressure derivatives of the bulk modulus $B'$ are also accurately described. The discrepancies between the theoretical results and experimental data shown in TABLE~\ref{tab:zerok} are due to the fact that experimental measurements are taken at non-zero temperatures. When the lattice constants are computed from NN-driven MD simulations at 298~K the NN lattice constant for the bcc structure (4.2816\AA) is very close to the experimentally obtained value of 4.2908\AA ~(TABLE~\ref{tab:zerok}). The fcc lattice constant obtained from MD simulations at $T=298$~K and $P=70$~GPa (3.6570\AA) is also in very good agreement with the experimentally measured value of 3.6292\AA ~(the fcc crystal is not stable at low pressure and, thus, we did not attempt to obtain its zero-pressure lattice constant).

\begin{table}
\caption{\label{tab:zerok}Structural and elastic properties of solid sodium phases.}
\begin{ruledtabular}
\begin{tabular}{lcccccc}
                     & \multicolumn{3}{c}{BCC} & \multicolumn{3}{c}{FCC} \\
                     \cline{2-4}\cline{5-7}
                     & DFT & NN & Exp.          & DFT & NN & Exp. \\
\hline
$a_0$ (\AA)          & 4.2008 & 4.2018 & 4.2908\footnotemark[1] & 5.2967 & 5.2972 & 5.4061\footnotemark[1] \\
$B_0$ (GPa)          & 7.63   & 7.59   & 6.31\footnotemark[1]   & 7.624  & 7.613  & 6.433\footnotemark[1]  \\
$B'$                 & 3.722  & 3.501  & 3.886\footnotemark[1]  & 3.71   & 3.82   & 3.83\footnotemark[1]   \\
$c_{11}$ (GPa)       & 8.72   & 8.70   & 8.57\footnotemark[2]   & 8.74   & 8.62   & \\
$c_{12}$ (GPa)       & 7.09   & 7.03   & 7.11\footnotemark[2]   & 7.07   & 7.11   & \\
$c_{44}$ (GPa)       & 6.11   & 6.42   & 5.87\footnotemark[2]   & 5.86   & 5.85   & \\
$c'$ (GPa)           & 0.82   & 0.83   & 0.73\footnotemark[2]   & 0.83   & 0.75   & \\
\end{tabular}
\end{ruledtabular}
\footnotetext[1]{X-ray diffraction study at $T=298$~K~\cite{a:hanfland}.}
\footnotetext[2]{Ultrasonic test at $T=80$~K~\cite{a:martinson}.}
\end{table}

\begin{figure}
\includegraphics*[width=8.5cm]{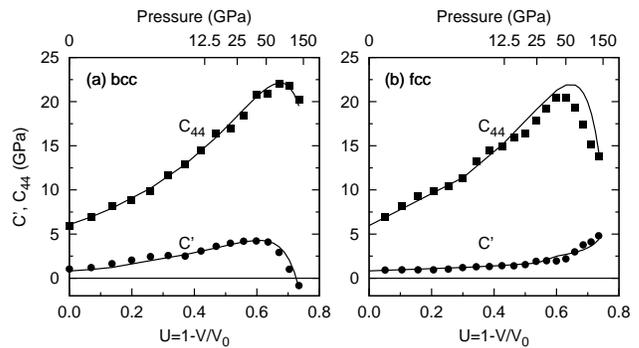}
\caption{\label{fig:elastic} Pressure dependence of the elastic coefficients. Points (lines) represent NN (DFT) data.}
\end{figure}

FIG.~\ref{fig:elastic} displays the calculated pressure dependence of selected elastic coefficients. A non-monotonic behavior of the tetragonal ($c'=\frac{c_{11}-c_{12}}{2}$) and trigonal ($c_{44}$) moduli obtained from \textit{ab initio} calculations is accurately reproduced with the NN. Elastic constant softening is a usual indication of a pressure-induced phase transformation. For example, it has been established that softening of the $c'$ modulus of bcc is connected with the bcc$\rightarrow$fcc structural transition along the tetragonal Bain path~\cite{a:katsnelson,a:xie}. It has also been suggested that the negative melting line in the sodium phase diagram could be related to the softening of elastic constants in crystal phases~\cite{a:pinsook}.

\begin{figure*}
\includegraphics*[width=17.5cm]{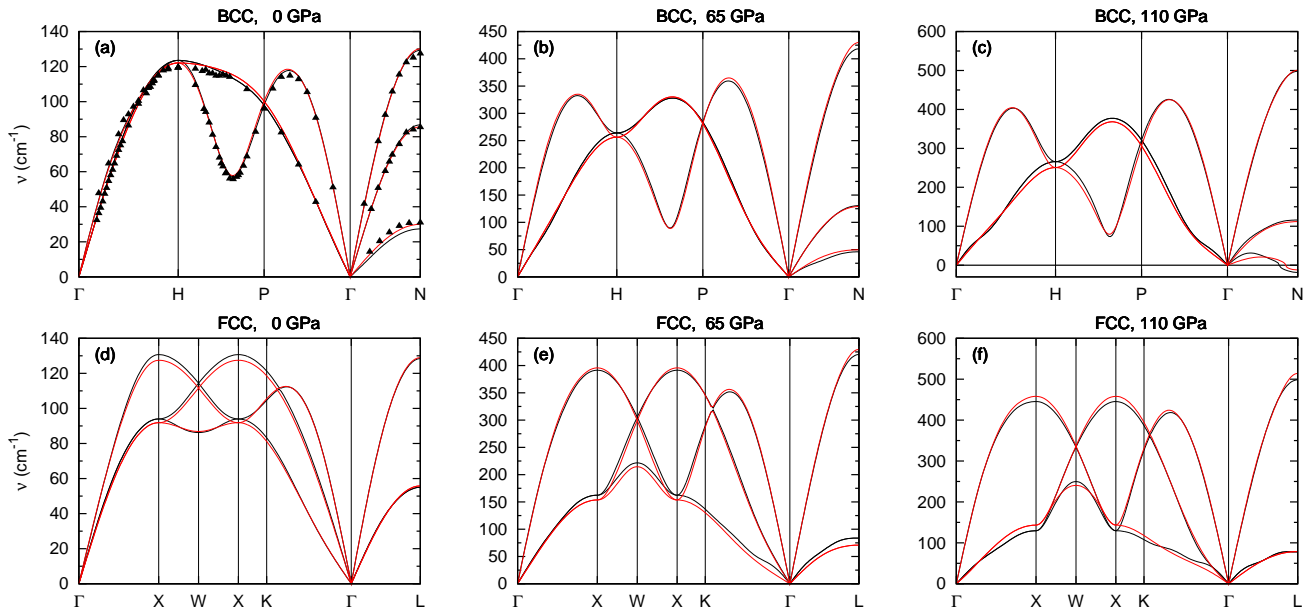}
\caption{\label{fig:phon} (color) Pressure dependence of the phonon dispersion curves. Red (black) lines correspond to NN (DFT) frequencies, triangles represent experimental data~\cite{a:expdisp}.}
\end{figure*}

The NN dispersion curves for the bcc and fcc structures are shown in FIG.~\ref{fig:phon} for a wide range of pressures. Remarkable agreement between the DFT and NN curves implies that the NN potential will provide results of DFT quality for all finite-temperature properties that are well described within the quasiharmonic approximation. The NN dispersion curves calculated for bcc at zero pressure (FIG.~\ref{fig:phon}a) are in closer agreement with experimental data than those predicted with the embedded atom potentials~\cite{a:hu10,a:hu} and the pseudopotential theory~\cite{a:magana}. We would like to point out that the pressure-induced transverse acoustic phonon softening along the [0$\xi\xi$]-direction near the zone center in bcc is accurately captured with the NN (FIG.~\ref{fig:phon}c). This softening is responsible for the instability to the tetragonal deformation (negative $c'$ in FIG.~\ref{fig:elastic}) and the bcc$\rightarrow$fcc transition mentioned above~\cite{a:xie}.

\begin{figure}
\includegraphics*[width=8.5cm]{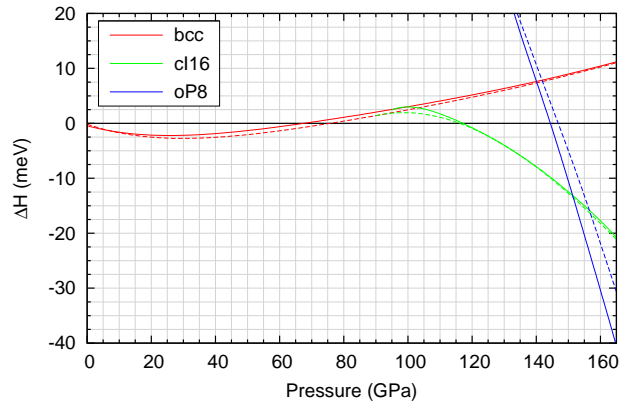}
\caption{\label{fig:dh} (color) Stability of crystal phases of sodium relative to the fcc phase. Solid (dashed) lines represent NN (DFT) results.}
\end{figure}

The calculated enthalpies of sodium crystal phases are shown in FIG.~\ref{fig:dh}. Despite small enthalpy differencies between the bcc and fcc phases ($\sim$2~meV/atom) the shape of the bcc NN enthalpy curve is in excellent agreement with the DFT data. The NN bcc$\rightarrow$fcc transition pressure (68~GPa) is only slightly lower than the corresponding DFT value (77~GPa) and in good agreement with experimental measurements (65~GPa at 298~K)~\cite{a:hanfland}.

\begin{figure}
\includegraphics*[width=8.5cm]{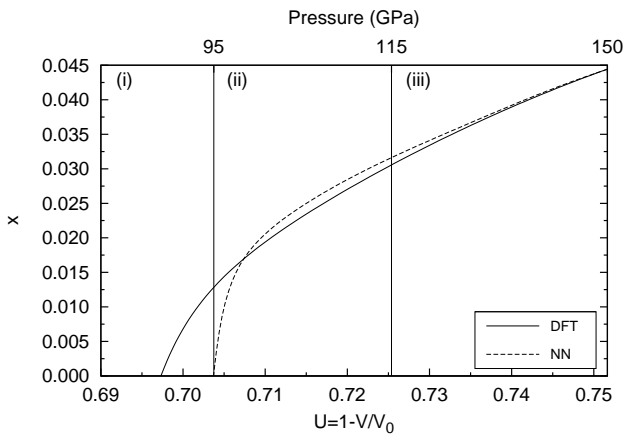}
\caption{\label{fig:ci16x} Distortion $x$ for cI16 as a function of compression $U$.}
\end{figure}

Experiments have shown that, in the pressure range from 105 to 117~GPa, the cI16 structure becomes the stable structure of sodium~\cite{a:syassen,a:mcmahon,a:gregor}. This structure can be represented as a 2$\times$2$\times$2 supercell of bcc with a small shift in the positions of the atoms along the space diagonal. This shift is described by a single internal parameter $x$ ($x=0$ gives the bcc structure). Our NN (DFT) calculations (FIG.~\ref{fig:dh} and FIG.~\ref{fig:ci16x}) show that cI16 becomes more stable than bcc at pressures above 95~GPa (90~GPa). The initial rapid growth of the distortion parameter $x$ in this pressure region [the dividing line between regions (i) and (ii)] is accompanied by only a small, on the order of 0.2~meV/atom, decrease of the enthalpy of cI16 relative to bcc. Such small energy differencies cannot be captured by the NN potential resulting in visible discrepancies between the $x$ values calculated with the NN and DFT. However, this discrepancy is only of minor importance in MD simulations since fcc is the most stable phase in this pressure range (i.e. the bcc$\rightarrow$cI16 transition is not observed experimentally). It is predicted by both NN and DFT that the cI16 strucutre of sodium becomes more stable than fcc above $\sim$117~GPa at zero temperature. This transition pressure is only slightly overestimated compared to 105~GPa experimentally observed for the fcc$\rightarrow$cI16 transition at $T=300$~K. Comparison of the NN and DFT predictions for the value of the distortion parameter $x$ demonstrates that the structure of the cI16 phase is well reproduced by the NN for the entire stability region of cI16 [region (iii) in FIG.~\ref{fig:ci16x}].

The NN and DFT calculations show that the oP8 structure becomes the stable phase of sodium at pressures above 160~GPa (the corresponding experimental transition pressure is 117~GPa). Thus, the NN potential accurately reproduces the experimentally observed sequence of stable sodium phases: bcc$\rightarrow$fcc$\rightarrow$cI16$\rightarrow$oP8. It predicts the transition pressures in very close agreement with DFT. 
The calculated (NN and DFT) zero-temperature transition pressures are above the experimentally observed room-temperature transition pressures with the discrepancy between theoretical and experimental $P_{tr}$ values increasing with pressure. The pressure of 150~GPa can currently be taken as the upper limit of the validity of the NN potential. Extension of the potential to higher pressures requires the inclusion of additional high-pressure structures into the fitting database.

\subsection{Liquid}

In this section, we test the performance of the NN potential for thermodynamic, structural, and dynamical properties of HPHT liquid sodium.

\begin{figure}
\includegraphics*[width=8.5cm]{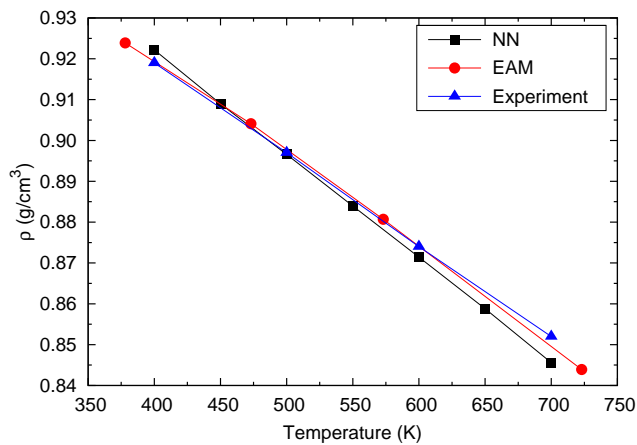}
\caption{\label{fig:liqthermal} (color online) Thermal expansion of sodium liquid at zero pressure. The EAM curve is from Ref.~\onlinecite{a:belashchenko}, the experimental curve is from Ref.~\onlinecite{a:kumari}.}
\end{figure}

\begin{figure}
\includegraphics[width=8.5cm]{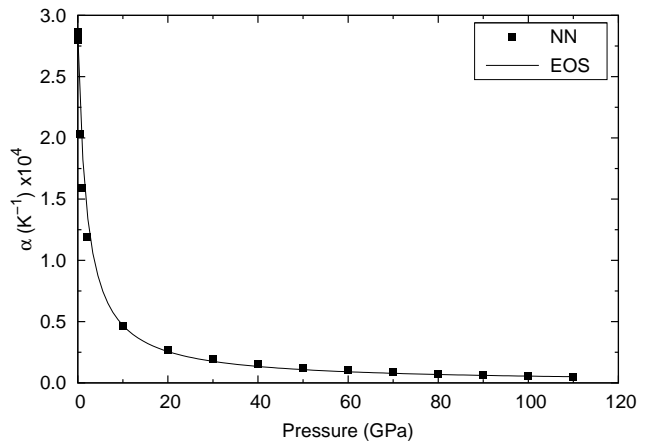}
\caption{\label{fig:thermalp} Pressure dependence of the volumetric thermal expansion coefficient of sodium liquid. The line is calculated using Eq.~\ref{eq:alphap}.}
\end{figure}

The temperature and pressure dependence of the density of liquid sodium is well described by the NN potential. FIG.~\ref{fig:liqthermal} shows that at zero pressure the NN density is in excellent agreement with the experimental and EAM data. The pressure dependence of the thermal expansion coefficient is shown in FIG.~\ref{fig:thermalp}. It is well described by the Murnaghan equation of state which postulates a linear dependence of the bulk modulus on pressure (the line in FIG.~\ref{fig:thermalp})~\cite{a:dass}
\begin{equation}\label{eq:alphap}
\alpha(P) = \alpha_0 \left( 1+\frac{B'_0}{B_0}P \right) ^{-1}.
\end{equation}
The zero pressure thermal expansion coefficient ($\alpha_0 = 2.861\cdot 10^{-4}$~K$^{-1}$), bulk modulus ($B_0 = 5.490$~GPa), and pressure derivative of the bulk modulus ($B'_0 = 2.792$) in FIG.~\ref{fig:thermalp} were determined from NN NPT simulations at 800~K. It is worth mentioning that the EAM of Belashchenko~\cite{a:belashchenko} predicts negative expansion coefficients at high pressures.

\begin{figure}
\includegraphics*[width=8.5cm]{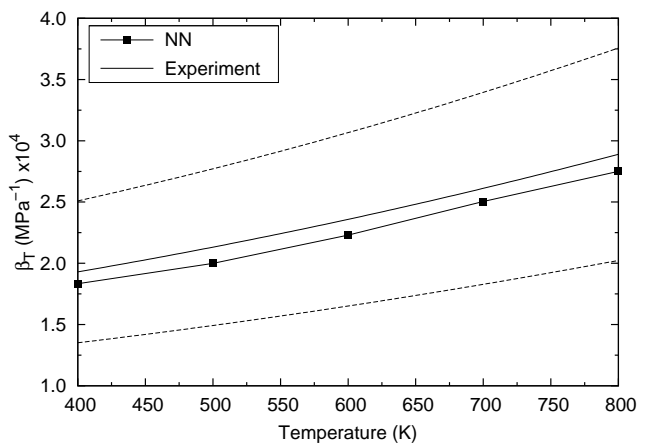}
\caption{\label{fig:compress} Isothermal compressibility of liquid sodium at zero pressure. The experimental curve is from Ref.~\onlinecite{a:expvisc}, the dashed lines show the uncertanty in the experimental data.}
\end{figure}

The isothermal compressibility of liquid sodium evaluated from the pressure dependence of the density is plotted in FIG.~\ref{fig:compress} for zero pressure. The NN predictions are in full agreement with the experimental results. The calculated isothermal compressibility at 100~GPa ($\beta_T = 3.75 \cdot 10^{-6}$~MPa$^{-1}$) is an order of magnitude smaller than the zero pressure value and does not change with temperature to any measurable degree. No experimental data is available for the isothermal compressibility at such a high pressure.

\begin{figure}
\includegraphics*[width=8.5cm]{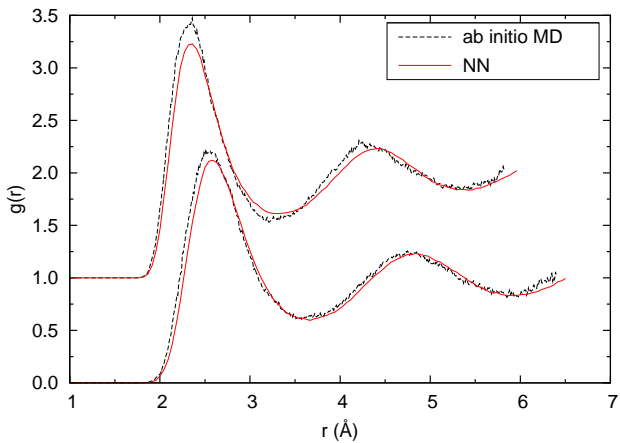}
\caption{\label{fig:rdf} (color online) Radial distribution functions of highly compressed liquid sodium. The lower curve is calculated at $\rho=2.8$~g/cm$^3$ and $T=1130$~K, the upper curve is obtained at $\rho=3.7$~g/cm$^3$ and $T=930$~K and is shifted up by 1. The \textit{ab initio} results are from Ref.~\onlinecite{a:yamane}}
\end{figure}

FIG.~\ref{fig:rdf} shows the radial distribution functions obtained with the NN potential at two state points ($\rho=2.8$ and $3.7$~g/cm$^3$ correspond to 50 and 110~GPa, respectively). Both of the functions are in good agreement with $g(r)$ calculated previously with \textit{ab initio} MD ($\rho=2.8$ and $3.7$~g/cm$^3$ correspond to 47 and 107~GPa, respectively)~\cite{a:yamane}. \textit{Ab initio} MD predicts slightly overstructured liquid compared to the NN simulations. The difference between the two methods is more noticeable at the higher pressure. This discrepancy is most likely due to inaccuracies (low plane wave cutoff and/or insufficient $k$-­point sampling) in the \textit{ab initio} simulations~\cite{a:yamane}. The \textit{ab initio} $g(r)$ in FIG.~\ref{fig:rdf} are calculated for a 54-atom system and $\Gamma$-point sampling of the Brillouin zone (54\textbf{k} points) while the NN is trained to reproduce high-quality \textit{ab initio} energies that correspond to 6000\textbf{k}-points.  

The transport properties of sodium liquid at zero pressure calculated with the NN potential are summarized in TABLE~\ref{tab:visc}. The experimental values of the self-diffusion coefficients are reproduced perfectly with the NN whereas the viscosity coefficients are systematically underestimated. The effect of temperature on the viscousity coefficients can be described with the exponential law:
\begin{equation}\label{eq:expn}
\eta = \eta_{\infty} \exp (E_a/RT),
\end{equation}
where $E_a$ is the activation energy of an elementary viscous flow process and $\eta_{\infty}$ is the pre-exponential factor that represents viscosity in the limit of zero activation barrier or infinite temperature. Fitting of the data to Eq.~\ref{eq:expn} shows that the activation energies are the same for the experimental and NN series (TABLE~\ref{tab:visc}). It is the pre-exponential factor, which includes the activation entropy of viscous flow, that is undestimated by $\sim$20\% in the NN calculations. Nevertheless, the 20\% error between the calculated and experimental values of $\eta$ is considered to be very small taking into account that the NN potential is derived only from \textit{ab initio} data~\cite{a:kuhnewater}.

\begin{table}
\caption{\label{tab:visc} Transport properties of liquid sodium at zero pressure.}
\begin{ruledtabular}
\begin{tabular}{ccccc}
T (K) & \multicolumn{2}{c}{D ($cm^2 \cdot sec^{-1}$) $\times 10^5$} & \multicolumn{2}{c}{$\eta$ ($Pa\cdot sec$) $\times 10^{4}$}\\
                & NN & Exp.\footnotemark[1] & NN & Exp.\footnotemark[2] \\
\hline
   407  &       5.681$\pm$0.034 &     5.44  &     4.709$\pm$0.275  & 5.810 \\
   448  &       7.307$\pm$0.033 &     7.16  &     3.917$\pm$0.152  & 4.935 \\
   490  &       9.070$\pm$0.001 &     9.05  &     3.513$\pm$0.003  & 4.282 \\
   500  &       9.499$\pm$0.169 &     9.52  &     3.388$\pm$0.601  & 4.152 \\
\hline
\hline
 & & & NN & Exp. \\
\hline
   & \multicolumn{2}{l}{$E_a$ ($J \cdot mol^{-1}$)}  & 6080 & 6105 \\
   & \multicolumn{2}{l}{$\eta_{\infty}$ ($Pa\cdot sec$) $\times 10^{4}$} & 0.777 & 0.957 \\
\end{tabular}
\end{ruledtabular}
\footnotetext[1]{Ref.~\onlinecite{a:meyerdiff}.}
\footnotetext[2]{Ref.~\onlinecite{a:russianvisc} as reported in Ref.~\onlinecite{a:expvisc}.}
\end{table}

\section{Summary and conclusions}

In summary, a NN potential for sodium has been created and tested to reproduce properties of HPHT crystal and liquid phases. The NN potential captures the experimentally observed sequence of pressure-induced solid-state phase transformations: bcc$\rightarrow$fcc$\rightarrow$cI16$\rightarrow$oP8. The transition pressures as well as structural, elastic, and vibrational properties of bcc, fcc, and cI16 phases predicted with the NN are in very close agreement with DFT. All calculated properties of bcc, fcc, and cI16 phases are in quantitative agreement with experimental data. The calculated (NN and DFT) zero-temperature transition pressures are above the experimentally observed room-temperature transition pressures. Further investigation of the transition pressure dependence on temperature is required to understand the origins of this discrepancy. The pressure of 150~GPa (on the NN and DFT scales) can currently be taken as the upper limit of the validity of the NN potential for crystal phases. The NN potential provides an \textit{ab initio} quality description of thermodynamics (density dependence on pressure and temperature), structural, and dynamical (diffusion and viscosity) properties of sodium liquid in the P--T region up to 120~GPa and 1200~K. 

The unique combination of accuracy and efficiency of the NN potential presented here will lead to dramatic enhancement of the quality of MD simulations and better understanding of the microscopic origins of complex behavior of HPHT phases of sodium. Detailed studies of the nature of the reentrant points on the melting curves in the sodium phase diagram and search for new crystal phases~\cite{a:rahman} are a few examples of useful follow-on developments of this work.

\begin{acknowledgments}
The authors would like to thank M. Ceriotti for his help with MD simulations that used a Langevin thermostat and barostat. JB is grateful for financial support by the FCI and the DFG. Our thanks are also due to the Swiss National Supercomputing Centre and High Performance Computing Group of ETH Z\"urich for computer time. 
\end{acknowledgments}

\bibliography{sodium_pot}

\end{document}